\documentclass[useAMS,usenatbib]{mn2e}

\usepackage{graphicx}
\usepackage{caption}
\usepackage{siunitx}
\usepackage[fleqn]{amsmath}
\usepackage{lipsum}
\usepackage{natbib}
\usepackage{amssymb}
\DeclareCaptionFormat{empty}{#1}

\title[Reconstructing the IMF of disc--bulge globular clusters]{Reconstructing the initial mass function of disc--bulge Galactic globular clusters from $N$--body simulations}

\author[Rossi \& Hurley]{L. J. Rossi$^{1}$\thanks{E-mail: lucarossi@swin.edu.au}, J. R. Hurley$^{1}$\\
$^{1}$ Centre for Astrophysics and Supercomputing, Swinburne University of Technology, Hawthorn, VIC 3122, Australia}

\begin{document}

\date{Accepted ----. Received ----; in original form ----}

\pagerange{\pageref{firstpage}--\pageref{lastpage}} \pubyear{2014}

\maketitle

\label{firstpage}

\begin{abstract}
We propose an evolutionary model to describe the dynamical evolution of star cluster systems in tidal fields, in which we calibrated the parametric equations  defining the model by running direct $N$--body simulations of star clusters with a wide range of initial masses and set of orbital parameters, living within the external tidal field generated by a disc--like galaxy. We derived a new method to solve numerically the evolutionary equations, allowing us to infer constraints on the mass of a star cluster from its age, present--day mass, orbital parameters and external gravitational potential. The result has been applied to the metal--rich subsample of Galactic globular clusters, being a good representation of a disc--bulge population. We reconstructed the initial mass function of these objects from the present--day mass function, finding that a lognormal distribution is well preserved during the evolution of the globular cluster system. The evolution of a power--law initial mass function has been evaluated, confirming that it transforms into a lognormal distribution of the cluster masses within an Hubble time. Our results are consistent with a formation scenario in which metal--rich Galactic globular clusters formed from giant molecular clouds in high--pressure regions during the early phases of the evolution of the Galactic disc and bulge.
\end{abstract}


\section{Introduction}

The long--lived nature of the globular clusters (GCs), the possible universality of the turnover of the globular cluster initial mass function (GCIMF) and the discovery of extragalactic globular cluster systems make these objects a powerful tool for inferring information on the physical conditions of the initial phases of galaxy evolution. In fact, the initial masses of globular clusters may be strictly correlated with features of the mass spectrum of the giant molecular clouds from which they have formed \citep{Parmentier07}. In this sense, the possibility that the globular clusters present day mass function (GCPDMF) may preserve an imprint of the initial mass function justifies any attempt to reconstruct their dynamical history.

GCs represent an ideal laboratory to test theories of stellar evolution and stellar dynamics \citep[e.g.][]{Meylan97}. In fact, since they are a very good approximation of ``single stellar population'' objects\footnote{Recent evidence of the presence of multiple populations in the color--magnitude diagram of several Galactic globular clusters indicates that this paradigm is no longer true \citep[e.g.][]{Piotto07}. However, in the context of the evolution of globular cluster systems in galaxies, the timescale involved in the generation of multiple stellar populations is order of magnitudes smaller than the timescale of the evolution of the globular cluster population.}, it is possible to derive precise ages and chemical composition through isochrone fitting and spectroscopic analysis of single stars. The globular clusters also represent an ideal target for studies related to galactic astrophysics \citep[e.g.][]{Brodie06}. The small dispersion of their relative ages and the bimodality often observed in the distribution of their colours, and hence of their metallicities, could mirror different physical conditions at the epoch of formation. Furthermore, the structural parameters characterizing a cluster could reflect the effect of the interaction of the cluster itself with the distribution of matter of the host galaxy.

Globular cluster systems (GCSs) have been recognized in a wide variety of environments, such as dwarf galaxies \citep{vandenBergh07}, irregular galaxies \citep{Piatti05}, early--type galaxies \citep{Pota13} and spiral galaxies \citep{Harris10}. They are also associated to mergers of gas--rich galaxies, which can lead to gravitationally driven increases in gas pressure that can in--turn trigger intense bursts of star and cluster formation \citep{Schweizer04}. The observed bimodality of the distribution of their colours can be interpreted in the context of a scenario in which GCs were formed in a dissipative process in the early Universe \citep[for a review see][]{Brodie06}. A possible scenario requires a mechanism to cut-off or truncate the formation of metal--poor GCs before the second phase of metal--rich GC formation occurred. Reionization could be the responsible process for this truncation \citep{Forbes97}. Another possible mechanism has been proposed by \cite{Tonini13}, directly related to the hierarchical galaxy assembly scenario. According to this model the metal-rich globular cluster subpopulation formed in the galaxy main progenitor, while the metal-poor
subpopulation has been accreted from satellites. The bimodality is observed also for Galactic GCs: the metal--poor GCs are usually associated to an extended inner halo population, while the metal--rich GCs belong to a bulge/thick disk population \citep{Zinn85}. In the Milky Way a third population of GCs orbiting in the halo has also been identified, associated with the accreted Sgr and Canis Major dwarf galaxies \citep{Forbes04}. 

Observational evidence suggests then that the physical conditions in which these system have been formed are different, and they can be associated to different phases of galaxy evolution. In particular, the initial mass function (IMF) of the GCSs could be strictly correlated to the mass spectrum of the giant molecular clouds (GMCs) from which they have been formed. Also, it could be possible that different populations of star clusters have different IMFs. One of the problems that we aim to address is related to the fact that the properties  of a globular cluster system will change in time. There are in fact several physical processes that can contribute to dissolve a star cluster. It is possible to separate them into \textit{internal} (stellar evolution and two--body relaxation) and \textit{external} processes (tidal disruption, disc/bulge shocking and dynamical friction). A realistic evolutionary model for globular clusters dependent on their orbital parameters and on the external gravitational potential of the host galaxy can allow us to infer information on the GCIMF and then on the physical conditions at the epoch of their formation.

In the past, several studies have been devoted to investigate the initial mass function of the globular clusters and its evolution. A pioneering approach was proposed by \cite{Vesperini97}, in which the authors derived a set of equations describing the evolution of the clusters orbiting within the external potential of a point--mass galaxy and applied those equations to predict the evolution of the mass function of the Galactic globular cluster system in a subsequent paper \citep{Vesperini98}.  A further analysis of the dissolution of globular clusters has been proposed by \cite{Baumgardt03}. In their work, the host galaxy has been modelled by a logarithmic potential, suitable to describe a dark matter halo. The authors calibrated the equations governing the evolution of GCs through a set of $N$--body simulations. The main limitation of these studies is that the gravitational potential of the external galaxy has been described with an extremely simple model. \cite{McLaughlin08} instead ignored the galactic gravitational potential and focused on cluster evaporation driven by internal dynamics only, following a simple analytic approximation of the mass--loss rate as function of the cluster density to show that the turnover mass and width of the GC present--day mass function (PDMF) is sensitive to this parameter. An alternative approach to derive the IMF of a GCS has been proposed by \cite{Parmentier07}, who developed a model to predict the globular cluster IMF from the mass spectrum of the progenitor GMCs, taking into account the expected star formation efficiency and the predicted fraction of stars bound to the cluster after the collapse of the proto--stellar cloud. A more general analysis has been proposed by \cite{Kravtsov05}, in which the authors studied the formation of globular clusters in a Milky Way--like galaxy using a high--resolution cosmological simulation. In a more recent work, \cite{Rieder13} presented a method to couple $N$--body star cluster simulations to a cosmological tidal field and compared the star cluster system in two Milky Way size haloes with a different accretion history. We also note the work of \cite{Alexander12} to present a prescription--based approach for following the evolution of various star cluster properties.

In this preliminary work we aim to develop an evolutionary model of globular clusters interacting with the static tidal field of a disc--like galaxy from direct $N$--body simulations performed with the GPU version of \texttt{NBODY6} \citep{Nitadori12}. The main improvement with respect to previous studies is that the present description of the external tidal field in \texttt{NBODY6} is more realistic. We have also been able to follow the evolution of a representative sample of clusters with an initial number of stars $N_0$ up to $10^5$. This has enabled us to present a method to solve numerically the evolutionary equations, obtaining information on the initial mass of the clusters from their age, observed mass and orbital parameters. We then applied this procedure to the bulge/thick disc subpopulation of Galactic globular clusters to predict the shape of the IMF for the metal--rich population and compared our results with previous works. We also followed the evolution of a truncated power--law IMF, evaluating the effects of the choice of its parameters on the evolution of a GCS mass distribution.

\section{Description of the models}

\subsection{Galactic mass model}
For the present work we assumed the galactic mass distribution model already implemented in \texttt{NBODY6} \citep{Aarseth03}. It consists of a point-mass bulge, a Miyamoto--Nagai disc \citep{Miyamoto75} and a logarithmic halo. In Cartesian right--handed Galactocentric coordinates:
\begin{equation}
\Phi_\mathrm{b} = -\dfrac{G M_\mathrm{b}}{\sqrt{x^2 + y^2 + z^2}}
\end{equation}
\begin{equation}
\Phi_\mathrm{d} = -\dfrac{G M_\mathrm{d}}{\{x^2 + y^2 + [a_\mathrm{d} + (b_\mathrm{d}^2 + z^2)^{1/2}]^2\}^{1/2}}
\end{equation}
\begin{equation}
\Phi_\mathrm{h} = -\dfrac{V_\mathrm{h}^2}{2}\log(a_\mathrm{h}^2 + x^2 + y^2 + z^2)
\end{equation}
where $\Phi_\mathrm{b}$, $\Phi_\mathrm{d}$ and $\Phi_\mathrm{h}$ are the gravitational potentials associated to the bulge, disc and halo respectively. In the present notation, $M_\mathrm{b}$ and $M_\mathrm{d}$ are the masses of the bulge and disc component, $a_\mathrm{d}$, $b_\mathrm{d}$ and $a_\mathrm{h}$ are scale lengths, $V_\mathrm{h}$ is the asymptotic value of the circular velocity curve and $G$ is the gravitational constant.  We also recall the relation between the circular rotation curve in the plane $z=0$ and the total gravitational potential
\begin{equation}
v_\mathrm{c}(R) = \sqrt{R\dfrac{d\Phi_\mathrm{tot}}{dR}}
\end{equation}
where $\Phi_\mathrm{tot} = \Phi_\mathrm{b} + \Phi_\mathrm{d} + \Phi_\mathrm{h}$ and $R^2=x^2 + y^2$. The values of the parameters have been chosen according to Model III in the work of \cite{Irrgang13} and they are summarized in Table \ref{tab:parameters}, while the parameters of the dark matter halo can be constrained by the velocity of the local standard of rest and by the distance of the Sun from the Galactic center:
\begin{equation}
V_\mathrm{h}=\left\{v_\mathrm{LSR}^2 - \dfrac{M_\mathrm{b}}{R_{\mathrm{gc},\odot}}-\dfrac{M_\mathrm{d}R_{\mathrm{gc},\odot}^2}	{[R_{\mathrm{gc},\odot}^2+(a_\mathrm{d}+b_\mathrm{d})^2]^{3/2}}\right\}^{1/2}
\end{equation}
\begin{equation}
a_\mathrm{h}^2=R_{\mathrm{gc},\odot}^2\dfrac{(v_\mathrm{LSR}^2-V_\mathrm{h}^2)}{V_\mathrm{h}^2}\;\;\;.
\end{equation}
We have assumed a Galactocentric distance of the Sun  $R_{\mathrm{gc},\odot} = 8.33 \; \mathrm{kpc}$ \citep{Gillessen09} and  a velocity of the local standard of rest $v_\mathrm{LSR}=239.7 \; \mathrm{km}\;\mathrm{s}^{-1}$ \citep[Model III]{Irrgang13}. 
\begin{table}
\centering
\begin{tabular}{l | l}
\hline
\hline
\\
$M_\mathrm{b}$  & $1.0\times10^{10} \; \mathrm{M}_\odot$\\ 
$M_\mathrm{d}$  & $7.2\times10^{10} \; \mathrm{M}_\odot$\\ 
$a_\mathrm{d}$  & $3.26 \; \mathrm{kpc}$\\ 
$b_\mathrm{d}$  & $0.29 \; \mathrm{kpc}$\\ 
$R_\odot$ & $8.33 \; \mathrm{kpc}$\\
$v_\mathrm{LSR}$ & $239.7 \; \mathrm{km} \; \mathrm{s}^{-1}$\\
\\
\hline
\hline
\end{tabular}
\caption{Parameters of the mass model.}
\label{tab:parameters}
\end{table} 

\subsection{Initial set up of the simulations}
We designed a wide set of $N$--body simulations of clusters with different initial masses and following different trajectories. In particular, we selected values of $N_0=1\times10^{4}, \;2\times10^{4}, \;3\times10^{4}, \;4\times10^{4}, \;5\times10^{4}, \;1\times10^{5}$ located on circular orbits at $R = 1, 2, 3, 4, 6$ and $8 \; \mathrm{kpc}$. In order to evaluate the effect of eccentricity and inclination from the Galactic plane, we ran simulations of clusters with different values of eccentricity in the range $0.0\leq e \leq 0.8$ and with values of inclination determined by applying a perturbation in the $z$ direction to initial conditions generating a circular planar orbit. We adopted the Kroupa stellar initial mass function \citep{Kroupa01} and a value of the tidal radius determined according to the external tidal field (see Appendix \ref{sec:appendix}). The stars are initially in virial equilibrium and distributed according to a Plummer sphere \citep{Plummer11}. We selected the average value of the metallicity for the Galactic disc--bulge GC population [Fe/H]=-0.5 \citep{Zinn85}, while the value of the length scale has been chosen in such a way that the clusters initially fill their tidal radius. The fraction of primordial binaries has been set equal to 5\% of the initial number of stars and the binary orbital setup has been chosen as described in \cite{Geller13}.

\section{Results}
We follow the distribution of the clusters until their dissolution, defined as the time at which only 300 stars remained gravitationally bounded. In fact, at small $N$ the results of simulations can become very noisy. For example, the presence (or not) of an energetic binary can result in an amplified effect on the evolution of the cluster at late times which could lead to variations in the result. Thus we chose this arbitrary cut--off to avoid small--$N$ fluctuations at very late stages. For the same reason the simulations which start with a small number of stars could be affected by statistical noise. We averaged the dissolution times of several simulations of the same cluster (primarily those with short dissolution times) to take this effect into account, but we found that the fluctuation in the dissolution time is smaller than the error associated to the fitting model. We ran all the simulations using \texttt{NBODY6} \citep{Aarseth03}, which also includes stellar and binary evolution as described in \cite{Hurley01}. 

\subsection{Dissolution time}
\begin{figure}
\centering
\includegraphics[scale=0.5]{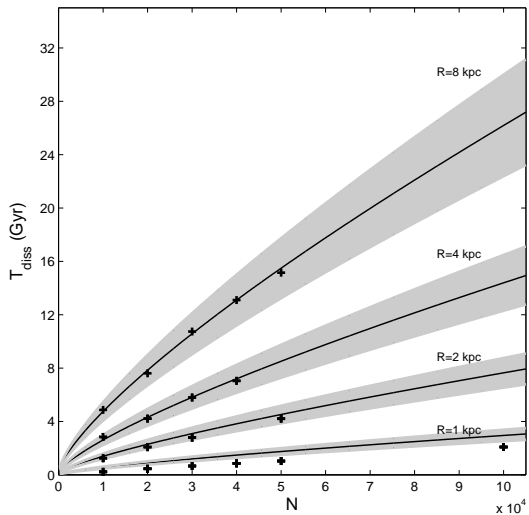} 
\caption{Fit of the dissolution times predicted by $N$--body simulations with the adopted model. The black crosses represent the results from the simulations, the black line is the best fit model and the shaded areas represent the uncertainties derived by a bootstrapping analysis.}
\label{fig:fit_tdiss}
\end{figure}

\begin{figure}
\centering
\includegraphics[scale=0.5]{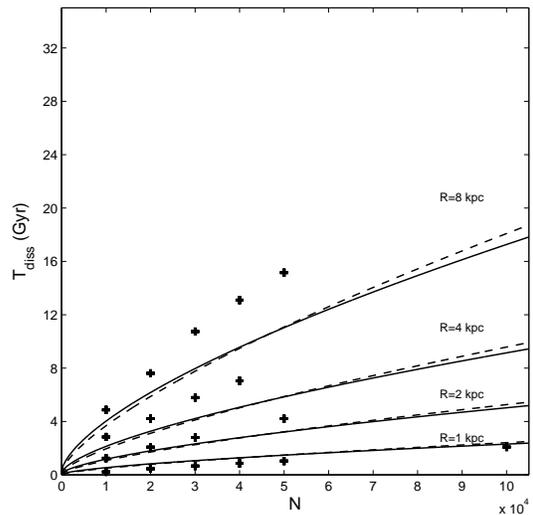} 
\caption{Prediction of the dissolution times from Baumgardt \& Makino (2003). The continuous line represents the result for an initial concentration parameter of the clusters equal to $W_0 = 5.0$, while the dashed line reproduces the result for $W_0 = 7.0$. Crosses represent our simulation results, as in Figure \ref{fig:fit_tdiss}.}.
\label{fig:BM_tdiss}
\end{figure}

According to \cite{Baumgardt03}, the dissolution time of a star cluster can be expressed in terms of the half--mass relaxation time and of the crossing time
\begin{equation}
t_\mathrm{diss} = k t_\mathrm{rh}^xt_\mathrm{cross}^{1-x}\;\;\;.
\label{eq:fundamental_tdiss}
\end{equation}
In particular
\begin{equation}
t_\mathrm{rh} \sim \dfrac{\sqrt{M_\mathrm{c}}r_\mathrm{h}^{3/2}}{\overline{m}\sqrt{G}\log(\gamma N)}
\end{equation}
is the half--mass relaxation time, where $M_\mathrm{c}$ is the total mass of the cluster, $r_\mathrm{h}$ is the half mass radius of the cluster, $\overline{m} = \sum _i m_i / N = M_\mathrm{c} / N$ is the mean mass of the stars, $N$ is the total number of stars, $\gamma$ is the Coulomb logarithm and $G$ is the gravitational constant. The second term in equation (\ref{eq:fundamental_tdiss}) is the crossing time
\begin{equation}
t_\mathrm{cross} \sim \dfrac{r_\mathrm{h}^{3/2}}{\sqrt{G M_\mathrm{c}}}\;\;\;.
\end{equation}
In fact, by definition, $t_\mathrm{cross} = R/v$ with $R \sim r_\mathrm{h}$. Assuming virial equilibrium
\begin{equation}
t_\mathrm{cross} \sim r_\mathrm{h} \sqrt{\dfrac{r_\mathrm{h}}{GM_c}} \sim \dfrac{r_\mathrm{h}^{3/2}}{\sqrt{G M_\mathrm{c}}}\;\;\;.
\end{equation}
Expanding the various terms in equation (\ref{eq:fundamental_tdiss}) we find
\begin{eqnarray}
t_\mathrm{diss} &=& k \left[ \dfrac{N r_\mathrm{h}^{3/2}}{\sqrt{G M_\mathrm{c}} \log(\gamma N)}\right]^x \left[ \dfrac{r_\mathrm{h}^{3/2}}{\sqrt{G M_\mathrm{c}}}\right]^{1-x}\nonumber\\
                &=& k \left[ \dfrac{N}{\log(\gamma N)}\right]^x \dfrac{r_\mathrm{h}^{3/2}}{\sqrt{G M_\mathrm{c}}}\;\;\;.
\end{eqnarray}
If we assume that all the radii defining the properties of a star cluster scale with its tidal radius (equation \ref{eq:general_rt}), then
\begin{equation}
t_\mathrm{diss} = k \left[ \dfrac{N}{\log(\gamma N)}\right]^x \left[ \dot{\theta}^2-\dfrac{d^2\Phi(R)}{d R^2}\right]^{-1/2}
\label{eq:tdiss}
\end{equation}
where $\dot{\theta}$ is the angular velocity of the cluster. For the specific case of a circular orbit
\begin{equation}
t_\mathrm{diss} = k \left[ \dfrac{N}{\log(\gamma N)}\right]^x \left[ \dfrac{1}{R}\dfrac{d\Phi(R)}{dR}-\dfrac{d^2\Phi(R)}{d R^2}\right]^{-1/2}\;\;\;.
\label{eq:tdiss_circular}
\end{equation}
The dissolution time of a cluster following a circular trajectory on the Galactic plane depends on the radius of its orbit and on its mass. We used the full set of $N$--body simulations to estimate the value of the coefficients $x$ and $k$, while their uncertainties have been estimated through a bootstrapping analysis. In Figure \ref{fig:fit_tdiss} we show the result for the best fitting model and related uncertainties from the fitting procedure.
Equation (\ref{eq:tdiss_circular}) can predict with good accuracy the dissolution time of clusters with different masses orbiting at different distances from the Galactic centre. The best values obtained for the model coefficients are $x = 0.88$, $k = 256.53$. The value of the $x$ parameter is comparable with the results from the analysis of \cite{Baumgardt03}, who found $x = 0.75-0.82$ with a weak dependence on the initial structure of the model clusters, although their predictions systematically underestimate the dissolution times of our model clusters for $R > 1 \; \mathrm{kpc}$ (see Figure \ref{fig:BM_tdiss}).
An explanation for such a difference in behaviour could be that the authors followed the evolution of clusters orbiting within a logarithmic gravitational potential, while the mass model assumed in this work includes also a representation for the bulge and for the disc. Also, the authors computed the value of the tidal radius valid in the case of orbits around a point mass, which could contribute to generate the observed inconsistencies.

\begin{figure}
\centering
\includegraphics[scale=0.5]{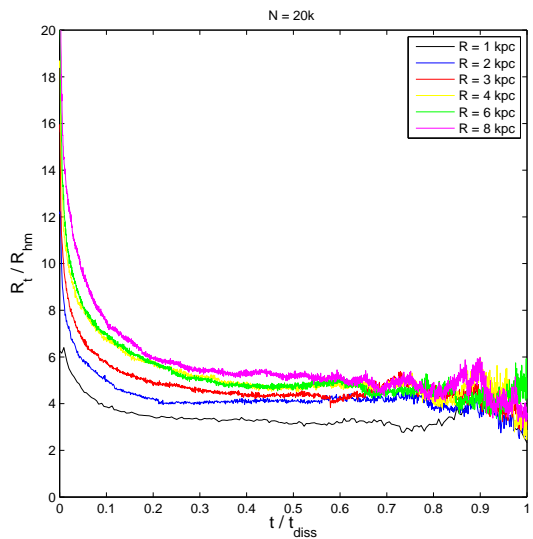}
\caption{Evolution of the ratio between half--mass radius and tidal radius for clusters with $N=20$k at different Galactocentric distances.}
\label{fig:rh_vs_rt}
\end{figure}

Although in general the model can predict with good accuracy the dissolution time of the simulated clusters, there are some inconsistencies for objects orbiting at 1 kpc from the Galactic centre. To determine the reason for this behaviour we looked for a breakdown in one of our fundamental assumptions, with the scaling of the half mass radius with the tidal radius being the most likely culprit. In Figure \ref{fig:rh_vs_rt} we show the problem for the clusters with $N = 2\times 10^4$. For all of our simulations the ratio between half mass radius and tidal radius eventually relaxes to a common value. For the clusters orbiting at 1 kpc, the value of this ratio is smaller than the value for outer orbits. This behaviour could be the combination of several effects, such as the strong tidal field in the inner kpc and the small time scales involved in the dynamical evolution of the inner clusters. Considering that for this set of simulations the bulge is described as a point mass, we shouldn't be too concerned about the very inner clusters, since an accurate description of their dynamical evolution requires more sophisticated models (see section \ref{sec:discussion}).

\subsection{Effects of orbital eccentricity}

\begin{figure}
\centering
\includegraphics[scale=0.5]{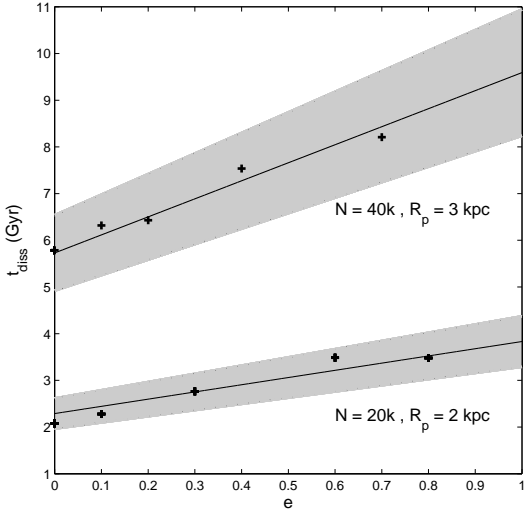}
\caption{Dissolution time as a function of the orbital eccentricity for clusters with the same initial mass and same perigalactic distance.}
\label{fig:e_tdiss}
\end{figure}

We found that the dissolution time scales linearly with the eccentricity of the orbit. The definition of eccentricity is $e = (R_\mathrm{a}-R_\mathrm{p})/(R_\mathrm{a}-R_\mathrm{p})$, where $R_\mathrm{a}$ and $R_\mathrm{p}$ are the apogalactic distance and the perigalactic distance respectively. The dissolution time as a function of the orbital eccentricity can be expressed in the form
\begin{equation}
t_\mathrm{diss}(e) = t_\mathrm{diss}(0)(1+\eta e)
\end{equation}
where $t_\mathrm{diss}(0)$ is the dissolution time of a cluster following a circular trajectory at the perigalactic distance and $e$ is the eccentricity. The best estimated value for the model coefficient is $\eta = 0.67$. Figure \ref{fig:e_tdiss} shows the results of our $N$--body simulations and the fit with a linear model.

\subsection{Effect of orbital inclination}
\label{subsec:inclination}
\begin{figure}
\centering
\includegraphics[scale=0.5]{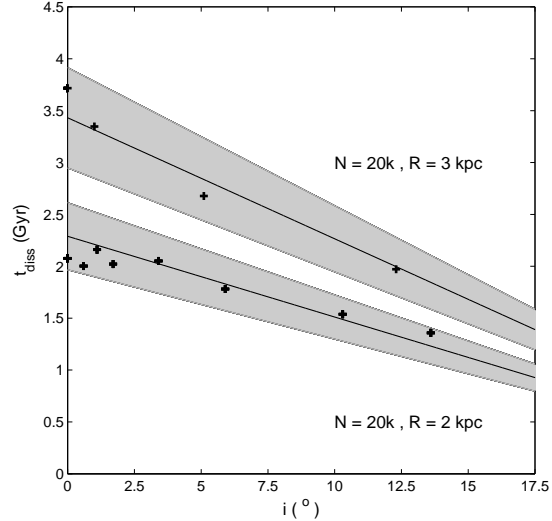}
\caption{Dependence of the dissolution time of the clusters on the orbit inclination.}
\label{fig:inc_tdiss}
\end{figure}

Similarly to the orbital eccentricity, we tried to evaluate the effect of inclined orbits on the lifetime of the clusters. This way we take into account the impact of disc shocking on the internal dynamics of star clusters. The results for a set of simulations for clusters following circular orbits on the galactic plane perturbed with an initial kick in the $z$ direction are shown in Figure \ref{fig:inc_tdiss}.
We found that the dissolution time scales in good approximation linearly with the inclination of the orbit, following a relation
\begin{equation}
t_\mathrm{diss}(i) = t_\mathrm{diss}(0)(1-\chi i)
\end{equation}
where $t_\mathrm{diss}(0)$ is the dissolution time for the cluster on a circular orbit on the galactic plane and $i$ is the inclination defined as $i = \mathrm{atan}(z/R)$, expressed in degrees. From a least squares interpolation, the best value obtained for the model parameter is $\chi = 0.03$, noting that we have only considered mild inclinations in this analysis. The reason for this is that for small inclinations the clusters describe a cylinder during their orbital evolution, while for the case of big excursions from the galactic plane the trajectory is more complex. This allows us to address the problem of describing the effect of disk shocking by adding a simple perturbing term to the equation predicting the dissolution time of clusters on circular planar orbits. With our model we can obtain realistic predictions for inclinations up to $i \simeq 20^{\circ}$, after which a more accurate description of disc shocking may be required \citep[e.g.][]{Gnedin97}.

\subsection{Mass loss rate} 
\begin{figure}
\centering
\includegraphics[scale=0.5]{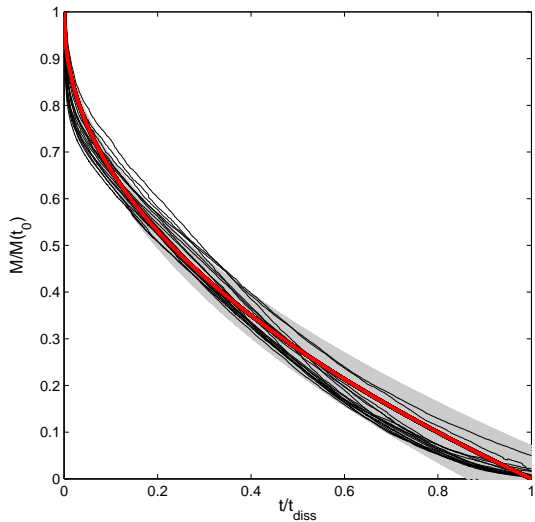}
\caption{Evolution of the clusters' mass normalized to the dissolution time and to the initial mass. The thick red line represents the best fit model. The uncertainties associated to the predicted value of the mass have been determined by the propagation of the error affecting our prediction of the dissolution time.}
\label{fig:mass_loss}
\end{figure}

In a recent study, \cite{Lamers10} looked in detail at the mass--loss rates and the mass evolution of star clusters. The authors identified four distinct mass--loss effects, namely mass loss by stellar evolution, the loss of stars induced by stellar evolution and the relaxation--driven mass loss before and after core collapse. We found a similar behaviour for the mass--loss rate (see Figure \ref{fig:mass_loss}), consisting of a steep initial decrease owing to stellar evolution, a linear trend dominated by relaxation and a final change likely due to post--core collapse  dynamics. We propose a simple model to interpolate the results of the simulations: the mass at the time $t$ of a cluster with an initial mass $M_\mathrm{c}(0)$ is modelled as a power-law of the form
\begin{equation}
M_\mathrm{c}(t) = M_\mathrm{c}(t_0)\left[ 1-\left(\dfrac{t}{t_\mathrm{diss}}\right)^{\xi}\right]
\label{eq:mass_loss_rate}
\end{equation}
with $t_0=0$. It is a slightly modified version of the model proposed by \cite{Baumgardt03} to take into account the non--constant mass--loss rate owing to stellar evolution effects. In Figure \ref{fig:mass_loss} we show the trend of the mass loss for our set of simulations normalised by the dissolution time and the initial mass. According to our results, the best value of the power--law shape is $\xi = 0.47$. Despite its simplicity, and considering the uncertainties in the dissolution time, this simple model can predict the mass of the cluster as function of time to a good approximation. 

\subsection{Dynamical friction and orbital decay}
\label{sec:Dyn_fric}
\begin{figure}
\centering
\begin{tabular}{c}
\includegraphics[scale=0.5]{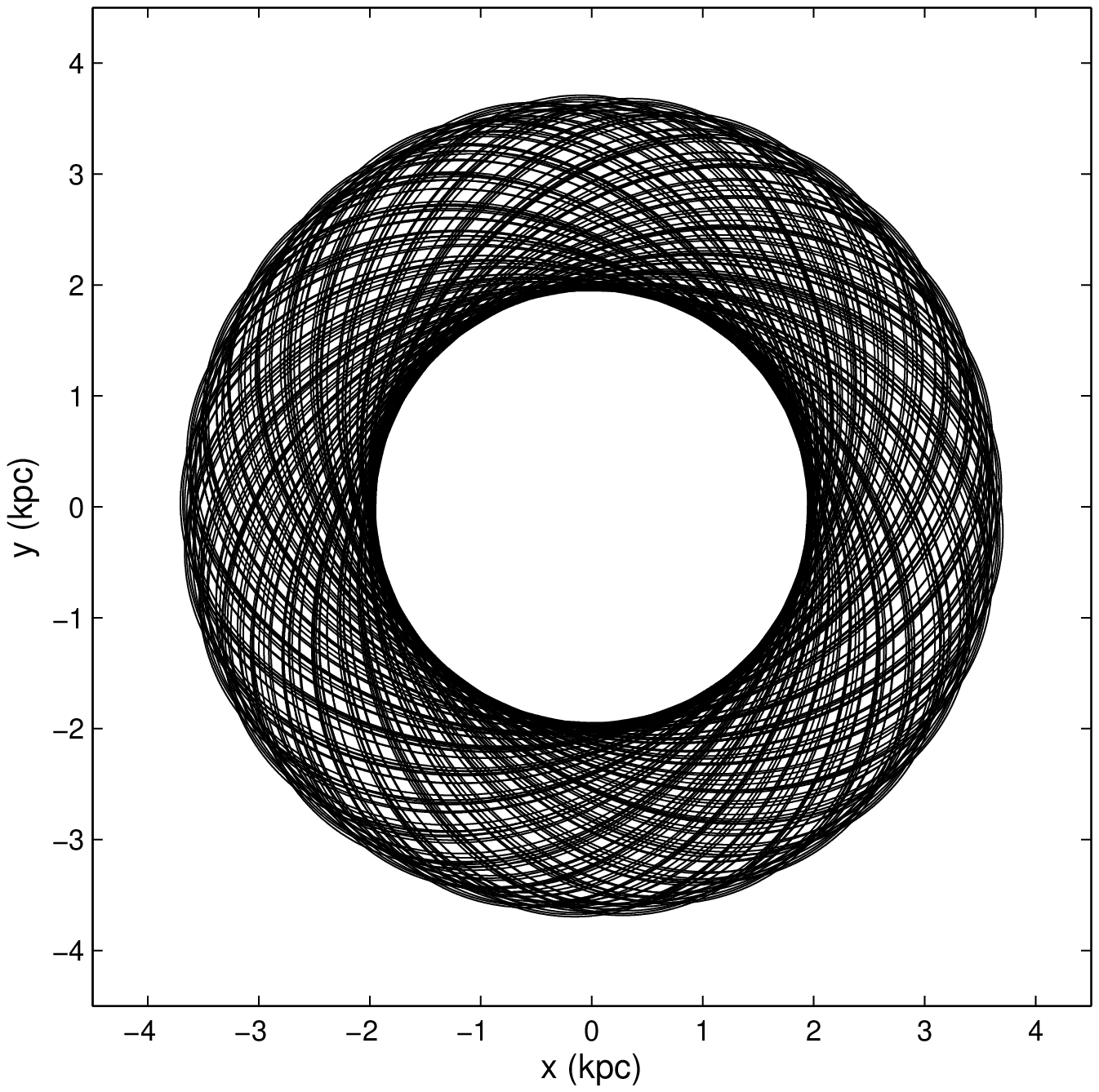} \\
\includegraphics[scale=0.5]{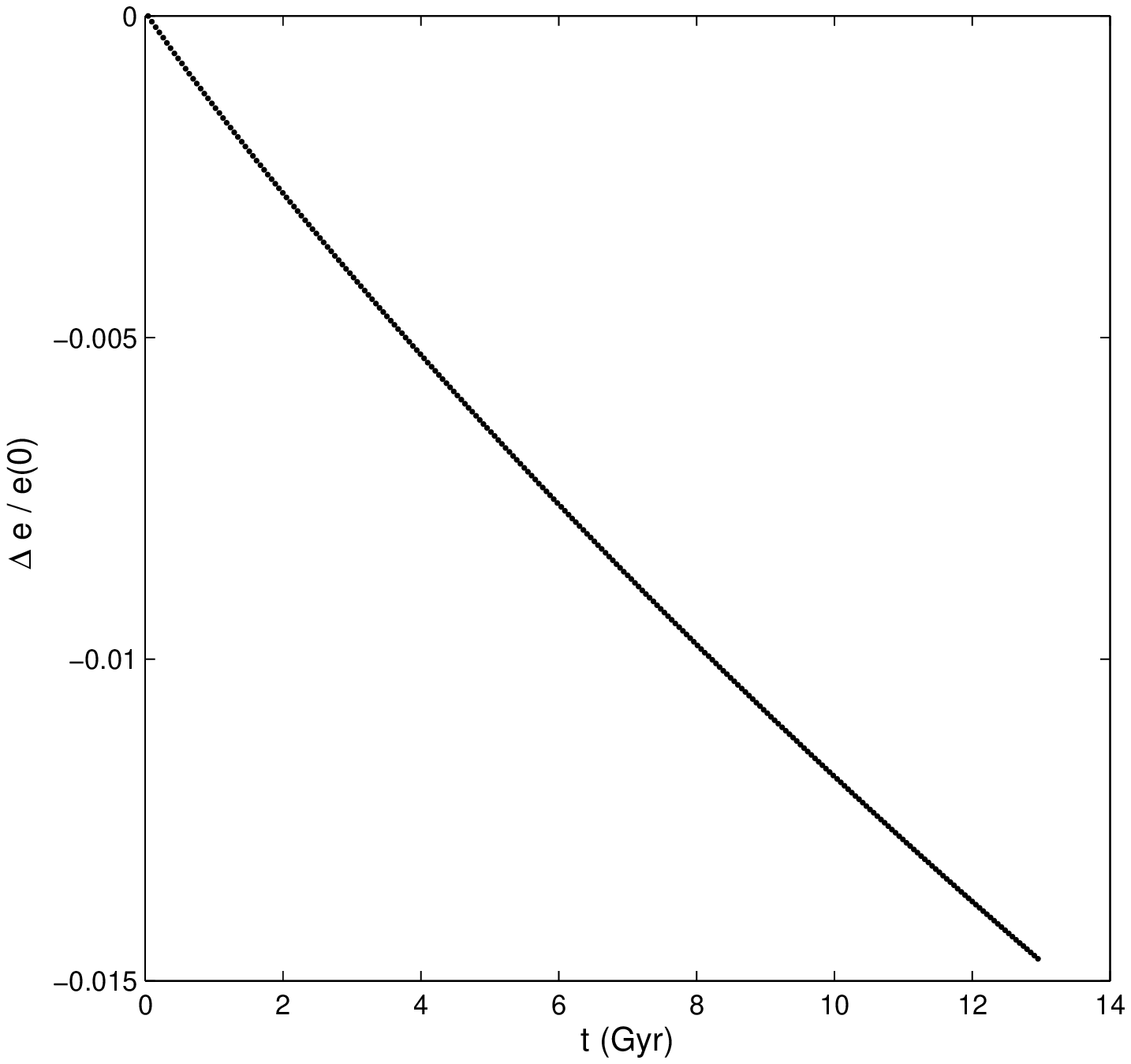}
\end{tabular}
\caption{Eccentricity decay and projection of the cluster orbit projected on the galactic plane. The decay of the orbit eccentricity is around the 1.5\% of the initial value.}
\label{fig:orb_decay}
\end{figure}

\begin{figure}
\centering
\includegraphics[scale=0.5]{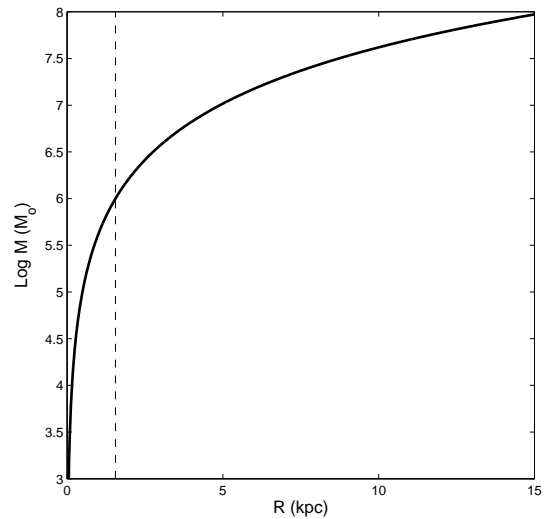}
\caption{Minimum initial mass of a cluster destroyed by dynamical friction within 12.8 Gyr as function of the Galactocentric distance. The dashed line represents the distance from the Galactic center for which the dynamical friction is relevant for an object with the mass of a typical globular cluster.}
\label{fig:Gnedin_df}
\end{figure}

We assumed that the clusters don't experience orbital decay during their evolution. Since in \texttt{NBODY6} the gravitational interaction with the parent galaxy is described in terms of analytic potentials, we want to obtain an estimate of the cumulative effect of the interaction of the clusters and the single field stars in order to justify this assumption. This phenomenon is well known as \textit{dynamical friction} and its effect is to decelerate the cluster in the direction opposite to its motion, causing a decay of the orbit.
The phenomenon of dynamical friction has been studied extensively in a series of papers by \cite{Chandra43}. Applying the fundamental results from kinetic theory, the author derived the famous equation
\begin{equation}
\dfrac{d \textbf{v}_\mathrm{M}}{dt} = -16 \pi^2 G^2M\mathrm{c} m_\mathrm{a} \log \Lambda\left[\int_0^{v_\mathrm{M}}dv_\mathrm{a}v_\mathrm{a}^2f(v_\mathrm{a})\right]\dfrac{\textbf{v}_\mathrm{M}}{v_\mathrm{M}^3}
\label{eq:Chandra_df}
\end{equation}
also known as the \textit{Chandrasekhar's  dynamical friction formula}. $M_\mathrm{c}$ is the mass of the target body (in this case the star cluster), $m_\mathrm{a}$ and $v_\mathrm{a}$ are the typical mass and velocity of a field star, $f(a)$ is the distribution function of the velocities of the field stars, $v_M$ is the velocity of target body and $\log \Lambda$ is the Coulomb logarithm
\begin{equation}
\log \Lambda = \log\left(\dfrac{b_\mathrm{max}}{b_\mathrm{min}}\right)\;\;\;.
\label{eq:coulomb_log}
\end{equation}
In this notation, $b_\mathrm{max}$ and $b_\mathrm{min}$ are the maximum and minimum impact parameters for gravitational encounters between the satellite cluster and the field stars in the parent galaxy. Following \cite{Hashimoto03}, $b_\mathrm{min}$ is set equal to the typical size of the cluster (we considered the tidal radius) and the cut--off radius is set as the distance of the clusters from the center of the parent galaxy, which varies during the orbital evolution. 
\cite{Fujii06}  demonstrated that this formulation of dynamical friction is not able to predict the exact results from $N$--body simulations because it doesn't consider additional effects such as the interaction between escaped particles and the clusters or the enhancement of dynamical friction by close escapers. Nevertheless, the formula provides a remarkably accurate description of the drag experienced by a rigid body orbiting in a stellar system if the value of the Coulomb logarithm is chosen appropriately. Following the approach proposed in \cite{Binney08}, it is possible to derive a simplified equation of the dynamical friction formula under the approximation of describing the density distribution of field stars by a singular isothermal sphere
\begin{equation}
\rho(r)= \dfrac{\Sigma^2}{2\pi G r^2} = \dfrac{v_\mathrm{c}^2}{4\pi G r^2}
\end{equation}
where $v_\mathrm{c}=\sqrt{2}\Sigma$ is the circular speed. The strongest justification of this assumption resides in the flatness observed in  many observed rotation curves. Since the distribution function for an isothermal sphere is Maxwellian, equation (\ref{eq:Chandra_df}) reduces to \citep[for details see][]{Binney08}
\begin{equation}
\dfrac{d\textbf{v}_\mathrm{M}}{dt}=-0.428 \log\Lambda \dfrac{G M_\mathrm{c}}{R^2}\;\;\;.
\label{eq:df_formula}
\end{equation}
The equation includes three time--dependent quantities, namely $\log \Lambda$, $M_\mathrm{c}$ and $R$. We have derived equations to predict the evolution of these quantities. The next step is to add a decelerating component to the equations of motion of a test particle orbiting within the gravitational potential implemented in \texttt{NBODY6} and integrate the orbit of a particle. We used an orbit integrator called \texttt{NIGO} (Numerical Integrator of Galactic Orbits) that we developed to support \texttt{NBODY6}. The results for a star cluster with an initial mass equal to $M_\mathrm{c}(t_0) = 1\times10^6 \; \mathrm{M}_\odot$ initially describing an eccentric orbit with $R_\mathrm{p} = 2 \; \mathrm{kpc}$ and $e = 0.3$ are shown in Figure \ref{fig:orb_decay}. The predicted value of the dissolution time for such a cluster is $t_\mathrm{diss}\sim 65 \; \mathrm{Gyr}$ and we followed its evolution for 13 Gyrs. The mass loss during the evolution and the change of the tidal radius of the cluster are taken into account.
Considering that the deceleration experienced by the target body is proportional to its mass and that we chose a relatively massive cluster, the obtained results represent an upper limit of the orbital decay that the models in this work could experience. Further considering that after 13 Gyr the decay of the orbital eccentricity is approximatively 1.5\% of the initial value, we can conclude that dynamical friction is negligible for the time scales we are considering. This conclusion is consistent with a result from the work of \cite{Gnedin14}, in which the authors derived the dynamical friction time--scale expressed in the form
\begin{equation}
t_\mathrm{df} = 0.45\;\mathrm{Gyr}\left( \dfrac{R}{\mathrm{kpc}}\right)^2 \left( \dfrac{V_c(R)}{\mathrm{km \; s^{-1}}}\right)
                \left( \dfrac{M(t)}{10^5\;M_\odot}\right)^{-1}f_\epsilon
\end{equation}
where $R$ is the initial radius of the orbit, $V_c(R)$ is the circular velocity curve, $M(t)$ is the mass of the cluster at a certain time $t$ and $f_\epsilon$ is the correction for eccentricity of cluster orbits (the authors assumed $f_\epsilon = 0.5$, we refer to the original paper for details). Figure \ref{fig:Gnedin_df} shows the limiting mass of a cluster destroyed by dynamical friction as function of the Galactocentric distance $r$ for a population of $t = 12.8 \; \mathrm{Gyr}$ (see section \ref{sec: reconstructed IMF}) and a constant value of the circular velocity equal to $V_c(R) = v_\mathrm{LSR} = 239.7 \; \mathrm{km} \; \mathrm{s}^{-1}$. We notice that the effect could be relevant for objects with a typical mass of a globular cluster ($10^4 \; M_\odot  \lesssim M_\mathrm{GC} \lesssim 10^6 \; M_\odot$) located in the inner regions of the Galaxy ($R \lesssim 1.5 \; \mathrm{kpc}$), while the mass limit increases rapidly above the typical mass range for greater distances from the Galactic center.

\section{Applications}
The evolution of the mass function of the Galactic globular cluster system has been investigated in two pioneering works by \cite{Vesperini97} and \cite{Vesperini98}. In the first paper the authors derived an analytic prediction of the evolution of the mass function basing their results on $N$--body simulations of star clusters with N = 4096, performed with NBODY4 \citep{Aarseth99}. In the second paper, the author investigated the evolution of the mass function of the Galactic globular cluster system adopting the derived analytical formulae. We note that the clusters simulated in these works evolved on circular orbits within the gravitational potential generated by a point-mass galaxy, while we propose a more accurate model of the galactic mass distribution and an analysis of the variation of the orbital elements on the intrinsic evolution of the star cluster.

\subsection{The mass function of Galactic metal--rich clusters}
\label{sec:IMF metal rich}

In this section we try to apply the results of our analysis of $N$--body simulations to real clusters. In particular, since the $N$--body simulations that we ran can be representative of a disc--like population of star clusters, we selected the metal--rich Galactic globular clusters. The main goal is to infer information on the initial mass function and on the initial radial mass distribution of these clusters from the observed present--day properties.

\subsubsection{The catalogue}
We selected metal--rich clusters from the Harris catalogue \citep[2010 version]{Harris96} by defining them as clusters with [Fe/H]$>$-1.0. We computed the  coordinates of the selected 49 clusters in the Galactocentric frame of reference assuming a distance from the Galactic centre equal to 8.33 kpc \citep{Gillessen09}. The catalogue contains information on the total visual magnitude of the clusters in V band, but not a direct estimation of their masses. We have to assume a certain value of the mass--to--light ratio in order to estimate the value of the cluster's mass from its luminosity.
     
\subsubsection{Mass--to--light ratio}
\label{sec:mass to light}
In a recent work, \cite{Bonatto12} proposed a new model to determine the mass of the clusters from their visual magnitude, and concluded that a power- law and a linear relation provide the best results in their fitting procedures. As such we have assumed a linear relation of the form
\begin{equation}
M/L_\mathrm{V} = a+ bL_\mathrm{V}
\end{equation} 
with $a = 0.43$ and $b = 0.20\times10^{-5}$ \citep[from][]{Bonatto12}. We converted the absolute $V$ magnitudes in $V$--band luminosities computing
\begin{equation}
L_\mathrm{V} = 10^{-0.4(M_\mathrm{V}-M_{\mathrm{V},\odot})}
\end{equation}
with $M_{\mathrm{V},\odot} = 4.83$. The mass of the clusters can then be expressed as
\begin{equation}
M= L_\mathrm{V}(a+ bL_\mathrm{V})\;\;\;.
\end{equation}

\subsubsection{Analysis of errors}
\begin{figure}
\centering
\includegraphics[scale=0.5]{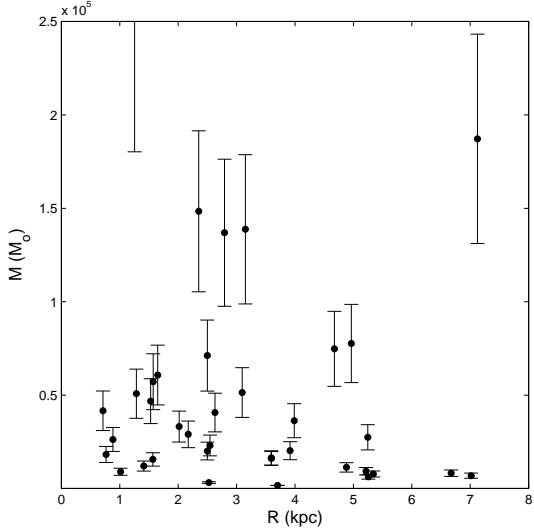}
\caption{Radial distribution of the Galactic metal--rich cluster masses from the Harris catalogue}
\label{fig:M_R}
\end{figure}

Since the catalogue doesn't provide the errors associated to magnitudes and distances, we estimated the uncertainties associated to our data points assuming that the heliocentric distance of a cluster is affected by a standard error equal to 10\% and propagating this error to the value of the mass.
The propagation of the error in the distance to the estimate of the absolute magnitude leads to

\begin{equation}
\sigma_{M_V} = \left| \dfrac{\partial M_\mathrm{V}}{\partial d}\right | \sigma_\mathrm{d} = \dfrac{5}{d \log 10} \dfrac{d}{10} = \dfrac{1}{2 \log 10} 
\end{equation}
using
\begin{equation}
M_\mathrm{V} = m_\mathrm{V} + 5 - 5\log_{10} d \;\;\;.
\end{equation}
The propagation to the  total $V$ luminosity leads to
\begin{equation}
\sigma_{L_\mathrm{V}} = \left| \dfrac{\partial L_\mathrm{V}}{\partial M_\mathrm{V}}\right | \sigma_{M_\mathrm{V}} = 0.4 \log10e^{(-0.4(M_\mathrm{V}-4.83)*\log 10)} \sigma_{M_\mathrm{V}}
\end{equation}
and, finally,
\begin{equation}
\sigma_M = \left| \dfrac{\partial M}{\partial L_\mathrm{V}}\right | \sigma_{L_\mathrm{V}} = (a + 2bL_V)\sigma_{L_\mathrm{V}}\;\;\;.
\end{equation}
Figure \ref{fig:M_R} shows the distribution of the masses of the metal--rich clusters and the associated uncertainties as function of their Galactocentric distance.

\subsubsection{Model fitting procedure}

\begin{figure}
\centering
\includegraphics[scale=0.5]{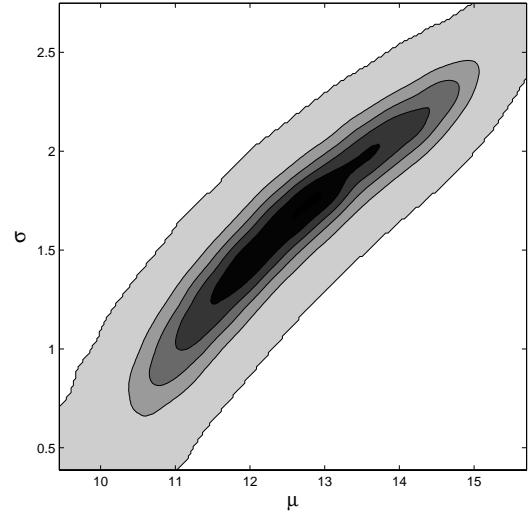} 
\caption{Contour plot of the distribution of the values of the model parameters in the lognormal function (Eq. \ref{eq:lognormal}) from the bootstrapping analysis.}
\label{fig:bootstrapping}
\end{figure}

We tried to fit the present day mass function of the metal--rich clusters with a lognormal distribution of the form
\begin{equation}
f(\mathcal{M}; K, \mu, \sigma) = \dfrac{K}{\mathcal{M}\sigma\sqrt{2\pi}}e^{-\dfrac{(\log \mathcal{M} -\mu)^2}{2\sigma^2}}
\label{eq:lognormal}
\end{equation}
where $K$ determines the amplitude of the distribution, $\mu$ the position of its peak and $\sigma$ its dispersion. We recall that for a lognormal distribution the position of the maximum is given by $M_\mathrm{max}=M_\mathrm{to}=e^{\mu+\sigma^2/2}$. We implemented a bootstrapping procedure to determine the best values of the fitting parameters and their uncertainties. The results of this  analysis is shown in Figure \ref{fig:bootstrapping}, where the distribution of the values in the plane ($\mu,\sigma$) has been plotted.

\begin{figure}
\centering
\includegraphics[scale=0.5]{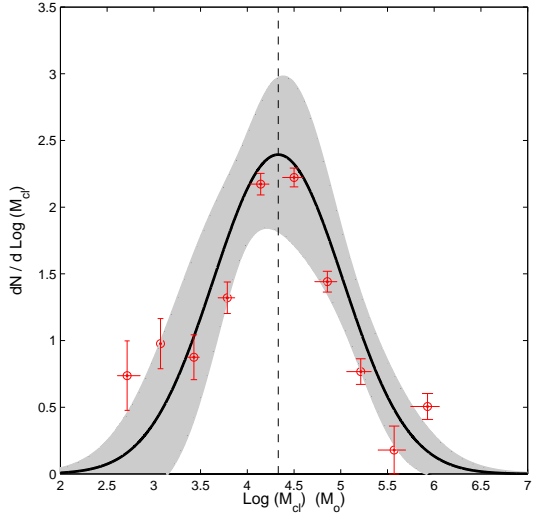} 
\caption{Present--day mass function of the metal--rich galactic globular clusters. The points represent the result form a logarithmic binning of the masses with associated uncertainties, the continuous line is the best fitting model and the shaded areas represent the uncertainties derived from a bootstrapping analysis. The dashed vertical line indicates the turnover mass.}
\label{fig:PDMF}
\end{figure}

We selected the mean value as the best estimate of the parameter and associated to this value the standard deviation of the data. Another major source of uncertainty is related to the small size of our sample (49 objects). In fact, the choice of the bin size affects the value of the parameters of the model, but we verified that the scatter in the parameter values owing to the choice of a reasonable bin size is included within the uncertainties estimated from the bootstrapping procedure. The best values derived for the fitting parameters are $K = 7.44\times 10^5$, $\mu = 12.55$ and $\sigma = 1.61$, while the value of the turnoff mass  for our metal--rich sample is $M_{\mathrm{to,PDMF}} = 2.14\times10^4 \; \mathrm{M}_\odot$. The final result of the model fitting procedure is the PDMF shown in Figure \ref{fig:PDMF}. On the vertical axis (number count) we included the statistical noise $\sigma_N = 1/\sqrt{N}$, where $N$ is the number of objects in each mass bin.

\subsubsection{Reconstructing the IMF of the surviving star clusters}
\label{sec: reconstructed IMF}
In the previous section we showed that the value of the mass of a cluster as function of its initial mass and orbital parameters can be expressed as
\begin{equation}
M(t) =  M(t_0)\left[ 1-\left(\dfrac{t}{t_\mathrm{diss}}\right)^{\xi}\right]
\end{equation}
where
\begin{equation}
t_\mathrm{diss} = \dfrac{k \left[ \dfrac{N}{\log(\gamma N)}\right]^x (1+\eta e)(1-\chi i)}{ \left[ \dfrac{1}{R}\dfrac{d\Phi(R)}{dR}-\dfrac{d^2\Phi(R)}{d R^2}\right]^{1/2}} \;\;\;.
\label{eq:evol_eq}
\end{equation}
The initial mass appears in the terms $M(t_0)$ and  $N \propto M_0$. The value of the proportionality constant depends on the initial distribution of the star masses in the clusters. For the specific case of a Kroupa stellar mass function that we are considering, its value is equal to $A = const = 0.6374$. We expressed the value of the initial mass as function of the age of the cluster $t$, the observed value of the mass $M(t)$ and the orbital elements. In particular
\begin{equation}
t_\mathrm{diss} = f(R,e,i)\left( \dfrac{M_0}{\log(\Lambda)}\right)^x
\end{equation}
where $N = AM_0$, $\Lambda = \gamma A M_0$ and
\begin{equation}
f(R,e,i) = \dfrac{k A^x (1+\eta e)(1-\chi i)}{ \left[ \dfrac{1}{R}\dfrac{d\Phi(R)}{dR}-\dfrac{d^2\Phi(R)}{d R^2}\right]^{1/2}}\;\;\;.
\end{equation} 
We can then expand
\begin{equation}
M_0 \left[1 - \left( \dfrac{t}{t_\mathrm{diss}}\right)^{\xi} \right] = M(t)
\end{equation}
\begin{equation}
M_0 \left[1 - \dfrac{t^{\xi}}{f(R,e,i)^{\xi}\left(\dfrac{M_0}{\log (\Lambda)}\right)^{\xi x}} \right] =  M(t)
\end{equation}
\begin{equation}
M_0 - M_0^{1-\xi x} (\log \Lambda)^{\xi x}\left(\dfrac{t}{f(R,e,i)}\right)^{\xi} =  M(t)\;\;\;.
\end{equation}
We can write
\begin{equation}
F(M_0) = M_0 - M_0^{1-\xi x} (\log \Lambda)^{\xi x}\left(\dfrac{t}{f(R,e,i)}\right)^{\xi} - M(t)
\end{equation}
which is, for every value of $R$, $e$, $i$ and $t$, of the form
\begin{equation}
g(x) = x + ax^{1-m}(\log bx)^m +c\;\;\;.
\end{equation}
A possible solution is to look for the zero  values of this function applying the Newton--Raphson method. In particular, we have to take the explicit first derivative of the function $g(x)$, i.e.
\begin{equation}
g'(x) = 1+a[(1-m)x^{-m}(\log bx)^m + mx^{-m}(\log bx)^{m-1}]
\end{equation}
and iterate until
\begin{equation}
x_{n+1} - x_n = -\dfrac{g(x_n)}{g'(x_n)}
\end{equation}
converges to an arbitrarily small value. This way we can extrapolate directly the initial value of the mass of the GCs.

\begin{figure}
\centering
\includegraphics[scale=0.5]{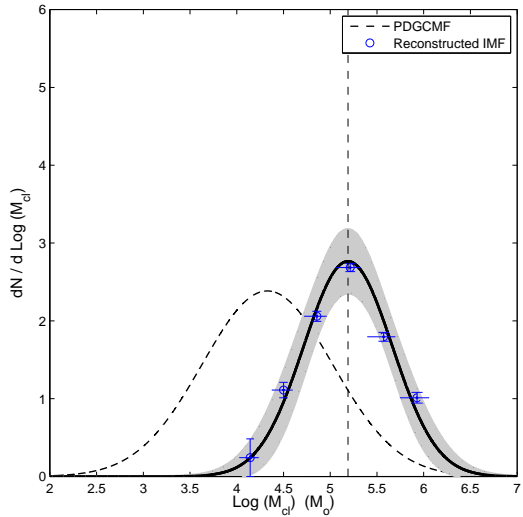}
\caption{Reconstruction of the IMF of the surviving metal--rich Galactic GCs in the general case of eccentric inclined orbits. The points represent the predicted initial mass of the 49 clusters in our sample binned logarithmically. The continuous line is a lognormal fit to the binned data and the thick dashed line is the PDMF from Figure \ref{fig:PDMF}.}
\label{fig:IMF_surviving}
\end{figure}

We applied this method to address the general problem of eccentric inclined orbits. We assigned to each cluster of our sample a random value of the eccentricity included in the range $0\leq e \leq 0.5$. The choice of this range of values has been suggested by the fact that the metal--rich GC population is associated with the Galactic bulge/thick--disc. In fact, according to \cite{Dinescu99}, clusters with thick--disc orbital characteristics are defined by inclination $\psi$ and eccentricity $e$ in the range ($0^{\circ} \leq i \lesssim 30^{\circ}$, $0 \leq e \lesssim 0.5$). We chose a representative value of the inclination determined as $i = \mathrm{atan}(z,R)$, where $z$ and $R^2 = x^2 +y^2$ are the present--day measured coordinates of the clusters. We assumed also that the clusters are coeval with $t = 12.8 \; \mathrm{Gyr}$ \citep{Marin09}. The results are shown in Figure \ref{fig:IMF_surviving} in terms of the predicted IMF (binned data points) and a lognormal fit to the binned data. The best values of the model parameters are $K = 1.95\times 10^6$, $\mu = 13.02$ and $\sigma = 1.05$, while the value of the turnoff mass is $M_{\mathrm{to,IMF}} = 1.51\times10^5 \; \mathrm{M}_\odot$.

\subsubsection{The missing population of dissolved star clusters}
\begin{figure}
\centering
\includegraphics[scale=0.5]{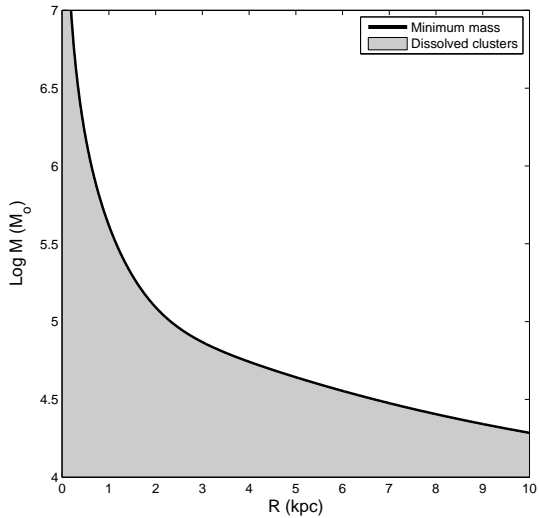}
\caption{Minimum mass of surviving star clusters as function of the Galactocentric distance for the case of planar circular orbits.}
\label{fig:M_lim}
\end{figure}

\begin{figure}
\centering
\includegraphics[scale=0.5]{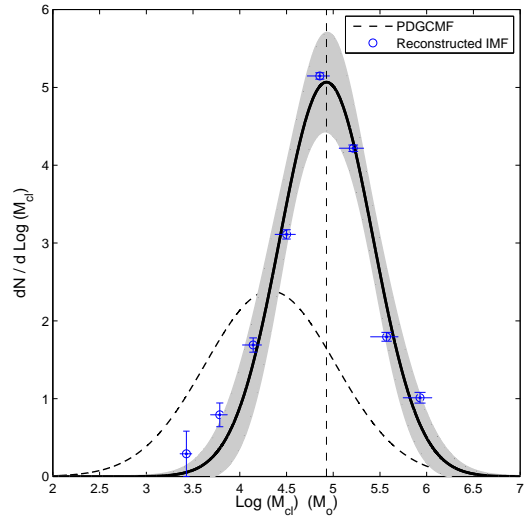}
\caption{Reconstructed IMF including a dissolved population of globular clusters. Symbols as in Figure \ref{fig:IMF_surviving}.}
\label{fig:IMF_complete}
\end{figure}

\cite{Vesperini98} predicted that the fraction of surviving clusters at the present epoch can be approximately half of the original cluster population. In the present section we propose a procedure to estimate the masses of the dissolved clusters in order to investigate how the IMF can be affected. Our model allows us to estimate the minimum mass $M_\mathrm{lim}$ of a surviving cluster (or, equivalently, $N_\mathrm{lim}$) as function of the Galactocentric distance. Assuming that the metal rich--clusters are coeval with  $t_\mathrm{mr} = 12.8 \; \mathrm{Gyr}$ the minimum mass of a surviving cluster can be determined under the condition
\begin{equation}
t_\mathrm{mr} = t_\mathrm{diss}
\end{equation}
which translates to
\begin{eqnarray}
N_\mathrm{lim}-\log(\gamma N_\mathrm{lim}) \left\{ \dfrac{t_\mathrm{mr}}{k(1+\eta e)(1-\chi i)}  \right.\nonumber\\
\times\left.\left[ \dfrac{1}{R}\dfrac{d\Phi(R)}{dR}-\dfrac{d^2\Phi(R)}{d R^2}\right]^{1/2}\right\}^{1/x} = 0
\label{eq:m_limit}
\end{eqnarray}
The solution of the last equation gives the minimum mass of a surviving star cluster as function of its Galactocentric distance. Figure \ref{fig:M_lim} shows the results of this analysis for the simple case of planar circular orbits ($e = 0$, $i = 0$).
A precise estimate of the properties of the dissolved star cluster population is hard to obtain, since all the information on this objects has been lost. We assumed to a first approximation that the number density of star clusters as function of the Galactocentric distance observed at the present day mirrors the initial one. This simplistic approximation could be justified considering the balance of two effects: the evidence that the observed density of star clusters decreases with the Galactocentric distance, and the prediction that disruption mechanisms are more efficient in the inner regions of the Galaxy. We created a synthetic population of star clusters equal in number to those observed today and characterized by the same radial number density distribution. Lacking any direct information, we assigned to each cluster a random value of eccentricity, inclination and mass, respectively, within the ranges of $0 \leq e \leq 0.5$, $0^{\circ} \leq i\leq 20^{\circ}$, $1\times10^3 \; M_\odot \leq M \leq M_\mathrm{lim}(R)$, where $M_\mathrm{lim}(R)$ has been determined according to equation (\ref{eq:m_limit}). We added these synthetic clusters to the IMF reconstructed from the surviving objects, obtaining the result shown in Figure \ref{fig:IMF_complete}. The best values of the model parameters are now $K = 2.37 \times 10^6$, $\mu = 12.66$ and $\sigma = 1.14$, while the value of the turnoff mass has decreased to $M_{\mathrm{to,IMF}} = 8.48\times10^4 \; \mathrm{M}_\odot$. While the shape of the high mass tail of the IMF has been little affected by the addition of the dissolved clusters, the low mass tail and the value of the turn-off mass are shifted towards lower values.

\subsection{Predicting the evolution of a truncated power-law IMF}

In the previous sections we showed that a lognormal IMF preserves its shape during the evolution of the globular cluster system. This evidence is in agreement with the results from the work of \cite{Vesperini98}, in which the author found that a Gaussian shape is well preserved during the entire evolution of the system. In this section we investigate the evolution of a generic power--law IMF. In this  particular case the distribution of the masses follows
\begin{equation}
dN = A M^{\alpha} dM
\end{equation}
or, equivalently,
\begin{equation}
\dfrac{dN}{d\log M} = A M^{\alpha +1}\;\;\;.
\end{equation}
The value of the constant $A$ is determined by the total number of objects $N_\mathrm{tot}$, by the shape of the mass distribution and  by the upper and lower limit of the mass range:
\begin{equation}
A = \dfrac{N_\mathrm{tot}}{\int_{M_\mathrm{low}}^{M_\mathrm{up}}M^{\alpha} dM}=\dfrac{N_\mathrm{tot}(1 +  \alpha)}{M_\mathrm{up}^{(1+\alpha)} - M_\mathrm{low}^{(1+\alpha)} }\;\;\;.
\end{equation}
In order to evaluated the effect of different values of the parameters defining the power--law IMF, we assumed that the Galactocentric distances of the clusters are randomly distributed in the range $0 \; \mathrm{kpc} \leq R_\mathrm{gc} \leq 15 \; \mathrm{kpc}$ (similar to the distance range of the observed metal--rich population) with eccentricity $0 \leq e \leq 0.5$ and inclination $0^{\circ} \leq i \leq 20^{\circ}$. In this analysis we neglected any correlation between the mass of the star cluster and its Galactocentric coordinates.
The effects of the choice of different mass limits and shapes of a power--law IMF are discussed. We note that in all the considered cases, \textit{an initial power--law mass function evolves into a lognormal mass function}.

\subsubsection{Effect of the power--law index of the IMF}
\begin{figure}
\centering
\includegraphics[scale=0.5]{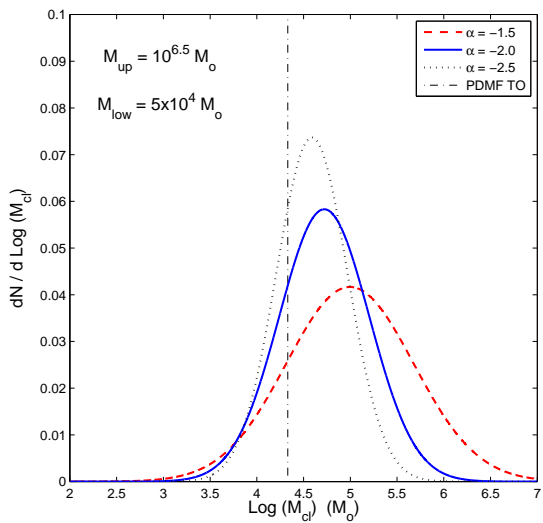}
\caption{Normalized evolved power--law MFs for different values of the slope $\alpha$. The dot--dash line represents the value of the observed turn-off mass of the PDMF.}
\label{fig:alpha_effect}
\end{figure}

In this section we studied the cases of $\alpha = -1.5, -2.0$ and $ -2.5$. Figure \ref{fig:alpha_effect}  shows the evolution of the normalized initial power law mass function for a globular cluster system $t = 12.8 \; \mathrm{Gyr}$ old for different values of the slope $\alpha$. The mass limits have been set equal to $M_\mathrm{low} = 5\times10^4 \; M_\odot$ and $M_\mathrm{up} = 10^{6.5} \; M_\odot$ (the effects of the choice of the mass limits are discussed below). Figure \ref{fig:alpha_effect} shows that both the value of the turnoff mass of an evolved power--law IMF and the dispersion $\sigma$ of the lognormal distribution increase for increasing steepness of the power--law IMF. 

\subsubsection{Effect of the low mass limit of the IMF}
\begin{figure}
\centering
\includegraphics[scale=0.5]{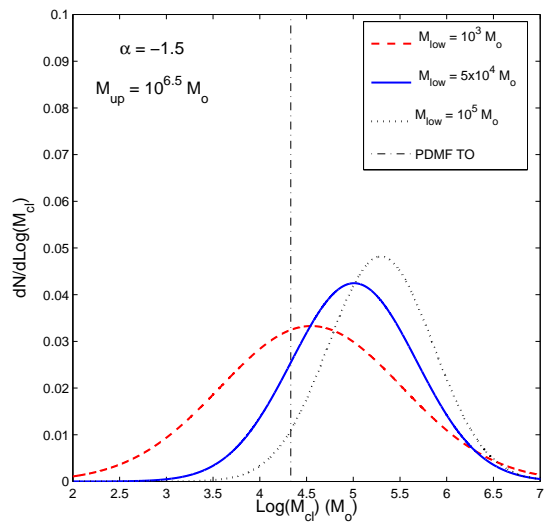}
\caption{Effect of the low mass limit on the evolution of a power--law IMF.}
\label{fig:low_mass_effect}
\end{figure}

As a test case, we followed the evolution of a power--law IMF with slope $\alpha= -1.5$, $M_\mathrm{up} = 10^{6.5} \; M_\odot$, $M_\mathrm{low} = 10^3,5\times10^4$ and $10^5 \; M_\odot$. Figure \ref{fig:low_mass_effect} shows the results of this analysis. We found that a smaller low--mass limit of a truncated power--law IMF results in a lower value of the turnoff mass and a higher value of the dispersion $\sigma$ of the evolved mass function.

\subsubsection{Effect of the high mass limit of the IMF}
\begin{figure}
\centering
\includegraphics[scale=0.5]{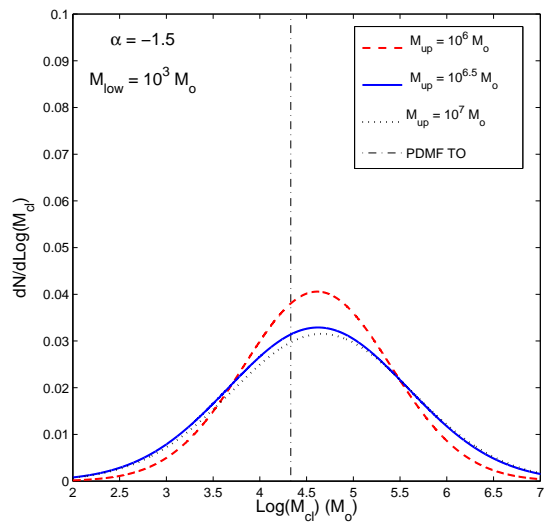}
\caption{Effect of the high mass limit on the evolution of a power--law IMF.}
\label{fig:high_mass_effect}
\end{figure}

We followed the evolution of mass functions with different values of $M_\mathrm{up}$ ($M_\mathrm{up} = 10^6 ,10^{6.5}$ and $10^7 \; M_\odot$) and shape $\alpha = -1.5$. Figure \ref{fig:high_mass_effect} shows the effects of the high--mass limit on the evolution of a power--law IMF. Comparing the results with the effects of the choice of the low--mass limit and of the slope of the IMF, the choice of $M_\mathrm{up}$ has only a minor impact on the properties of the evolved mass function. However, we notice that the dispersion of the evolved mass function mildly increases for increasing values of $M_\mathrm{high}$, while the value of the turnoff mass is almost unaffected.\\

\subsubsection{The case of the Milky Way metal--rich GCs}
\begin{figure}
\centering
\includegraphics[scale=0.5]{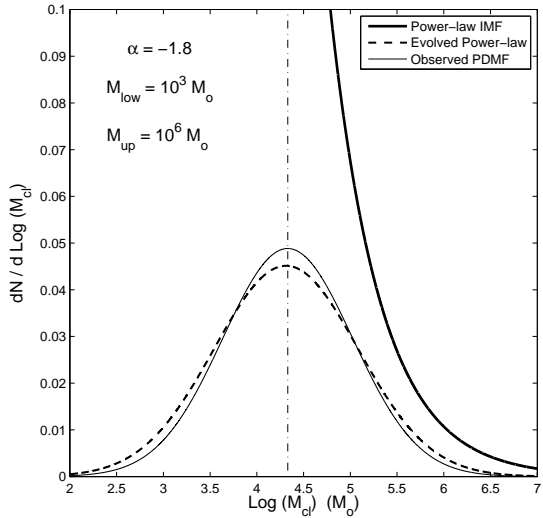}
\caption{Power--law IMF evolving into a Milky--Way like GCPDMF of the metal--rich population.}
\label{fig:MW_PL}
\end{figure}

Combining the effects of the choice of the parameters defining a power--law IMF, we found the the values of $\alpha = -1.8$, $M_\mathrm{up} = 10^{6} \; M_\odot$ snf $M_\mathrm{low} = 10^{3} \; M_\odot$ evolve into a mass function comparable with the PDMF of the metal--rich GCs population. The results are shown in Figure \ref{fig:MW_PL}.
This is a particularly interesting case, since a power--law with these characteristics is similar to the one describing the distribution of the masses of giant molecular clouds in the local Universe \citep[][we refer to section \ref{sec:discussion} for details]{Rosolowsky05}. 

\subsection{Reconstructing the radial distribution of the surviving clusters masses}
\label{sec:radial_mass_distribution}

\begin{figure}
\centering
\includegraphics[scale=0.5]{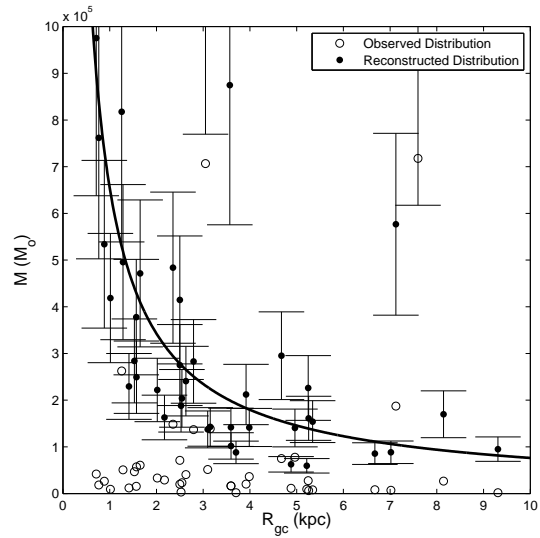}
\caption{Observed and reconstructed radial distribution of the surviving clusters masses, with the latter overlaid to a power--law regression curve.}
\label{fig:PDMD}
\end{figure}

The present approach allows us to reconstruct also the initial radial distribution of the surviving cluster masses, reminding the reader that we are focusing on the metal--rich population for an illustrative purpose. In Figure \ref{fig:PDMD} we show the observed distribution of the cluster masses as a function of their Galactocentric distance and the reconstructed distribution (with the assumption that the position has not changed with time: see Section \ref{sec:Dyn_fric}). We note that the observed distribution of the masses doesn't show a clear correlation with the cluster distance from the Galactic centre, while the predicted initial radial profile follows a power--law distribution of the form
\begin{equation}
M(R) \propto R^\gamma
\end{equation}
with $\gamma = -0.92$ from a non--linear least squares interpolation.

\subsection{The choice of the $M/L$ ratio}

The position of the turnoff observed in the PDMF depends on the value assumed for the mass--to--light ratio for the star clusters. In \cite{Mandushev91} the authors calibrated the mass--to--light ratio for Galactic globular clusters from the dynamical mass of a sample of clusters with reliable central velocity dispersion using the King formula (see the original paper for details):
\begin{equation}
M_\mathrm{c} = 167r_\mathrm{c}\mu_\mathrm{d}\sigma_0^2
\end{equation} 
where $M_\mathrm{c}$ is the mass of the cluster in solar masses, $r_\mathrm{c}$ is the core radius of the cluster in parsecs, $\mu_\mathrm{d}$ is the dimensionless mass of the model and $\sigma_0$ is the central velocity dispersion expressed in km s$^{-1}$. The derived relationship between the mass and absolute magnitude is
\begin{equation}
\log(M_\mathrm{c}/M_\odot) = -0.456 M_\mathrm{v} + 1.64 \;\;\;.
\end{equation}
We selected the same subsample of star clusters from the \cite{Mandushev91} catalogue that we previously identified as disc--bulge population objects and computed their PDMF. This resulted in a turnover of the PDMF corresponding to $M_{\mathrm{to,PDMF}} = 6.11\times10^4 \; \mathrm{M}_\odot$ which is higher than the value derived using the \cite{Bonatto12} $M/L$ ratio ($M_{\mathrm{to,PDMF}} = 2.14\times10^4 \; \mathrm{M}_\odot$). We found that the magnitude of the variation of the turnover mass of the IMF of surviving clusters derived assuming the value of the masses from \cite{Mandushev91} is comparable to the variation of the turnover of the PDMF ($\Delta M_{\mathrm{to,PDMF}} \simeq 3.0 \times10^4 \; \mathrm{M}_\odot$).

\section{Discussion and Conclusions}
\label{sec:discussion}

We have presented a method to simulate the evolution of star cluster systems in tidal fields, calibrating the evolutionary equations from the results of direct $N$--body simulations performed with \texttt{NBODY6}. 
Since our model at this time is suitable to simulate the dynamical evolution of clusters orbiting on mildly inclined orbits to the Galactic plane, we applied the results of the simulations to the Galactic metal--rich globular clusters, being a good representation of a bulge/thick disc population. We found that the IMF predicted from a direct solution of the evolutionary equations applied to the surviving clusters is well reproduced by a lognormal distribution of the cluster masses. In particular, the predicted initial value of the turnoff mass is $M_{\mathrm{to},\mathrm{IMF}} = 1.78\times10^5 \; \mathrm{M}_\odot$, while the present--day value is $M_{\mathrm{to},\mathrm{PDMF}} = 2.32\times10^4 \; \mathrm{M}_\odot$. We added a synthetic population of dissolved star clusters to the IMF of the surviving star clusters in order to evaluate the effect of the inclusion of a missing population. We found that a lognormal function well describes the distribution of the masses, although the low mass tail and the value of the turn-off mass are slightly shifted towards lower values.  In this sense, our results agree with the work of \cite{Vesperini98}, in which the author concluded that the lognormal distribution represents an ``equilibrium'' GCMF, able to preserve its shape during the evolution of the star cluster system.

We also followed the evolution of a truncated power--law IMF, finding that a slope $\alpha = -1.8$ and a mass range $1\times10^3 \; \mathrm{M}_\odot \leq M \leq 1\times10^6 \; \mathrm{M}_\odot$ reduces to a lognormal distribution with a turnoff mass and dispersion similar to the one observed for the metal--rich Galactic GCs. We evaluated the effects of the variation of the parameters defining the IMF (the slope $\alpha$, the low--mass truncation $M_\mathrm{low}$ and the high--mass truncation $M_\mathrm{high}$). According to our results, the value of the turnoff mass increases with the steepness of the power--law, decreases for decreasing values of $M_\mathrm{low}$ and is unaffected by the choice of $M_\mathrm{high}$. Also, the dispersion of the evolved mass function is greater for steeper power--laws and smaller values of $M_\mathrm{low}$, while in this case lower values of $M_\mathrm{high}$ result in smaller dispersions. There is a ``degeneracy'' in the globular cluster IMF, in the sense that both an initial lognormal MF and a truncated power--law MF can result in the observed bell--shaped distribution. Furthermore, the power--law IMF that evolves into a mass function observed for the metal--rich population is particularly interesting, being similar to the mass spectrum of the GMCs observed in the local group \citep{Rosolowsky05}. Also, according to \cite{Elmegreen96} the average value of the shape of the power--law mass of the GMCs in the local group is $\alpha = -1.8$. These authors also predicted that the bound cluster mass function should be similar to the progenitor cloud mass function, in good agreement with our result. On the other hand, considering that only a small fraction of the mass of a GMC can be converted into a bound star cluster, which will eventually expel the residual star forming gas due to supernova activity, we expect that the IMF of the globulars is different from the mass functions of their progenitors. This problem has been addressed by \cite{Parmentier07}, in which the authors proposed a relation 
\begin{equation}
M_\mathrm{c} = F_\mathrm{bound} \epsilon M_\mathrm{GMC}
\end{equation}
where $M_\mathrm{c}$ is the initial mass of the star cluster, $M_\mathrm{GMC}$ is the mass of the progenitor GMC, $\epsilon$ is the fraction of the gas mass converted into stars and $F_\mathrm{bound}$ is the fraction of stars bounded to the cluster after their formation. They found that the turnover of the GCIMF has a strong dependence on the lower mass limit of the progenitor GMCs, and hence a possible universality of the turnover mass would originate from a common value among galaxies for the lower mass truncation. According to the model proposed by \cite{Parmentier07}, the value of the turnoff of our reconstructed IMF (surviving + estimated dissolved clusters) is consistent with a lower limit for the progenitors $M_\mathrm{low} = 1\times10^5\;\mathrm{M}_\odot$. However,when comparing our results, we have to be aware that the GMCs are transient features of the interstellar medium, and it is not clear if the mass spectrum observed in the local Universe has changed from the epoch of cluster formation, mirroring the evolution of the galactic environment.

We evaluated also the evolution of the total mass of the metal--rich clusters system. According to our results, the present--day total mass is $M_\mathrm{PD} = 4.27\times10^6 \; M_\odot$. The total mass of the reconstructed IMF described by a lognormal distribution (surviving + estimated dissolved clusters) is $M_\mathrm{I,LN} = 1.35\times 10^7\; M_\odot$, while the total mass of the power--law IMF that evolves into the observed PDMF is   $M_\mathrm{I,PL} = 1.50\times 10^7\; M_\odot$. The total masses of the two different representations are comparable and represent a small fraction of the total mass of the disc--bulge component ($M_\mathrm{Disc+Bulge} = 8.2\times 10^{10}\; M_\odot$, according to the adopted Galactic mass model).

With the present approach we have been able to reconstruct the initial radial distribution of the masses of the surviving star clusters, finding that a power--law with slope $\gamma = -0.92$ interpolates the data points to a good approximation. The reconstructed profile is consistent with a scenario in which globular clusters of all ages preferentially form in high-pressure regions \citep{Elmegreen97}. This could be in good agreement with a formation mechanism in which the bulge formed from massive clumps of the Galactic disc, which spiralled to the center of the Galaxy and merge to form the central spheroid in a strong starburst \citep{Immeli04}. However, as shown in Figure \ref{fig:M_lim}, we may be missing information on clusters of lower masses that have already been dissolved.

\begin{figure}
\centering
\includegraphics[scale=0.5]{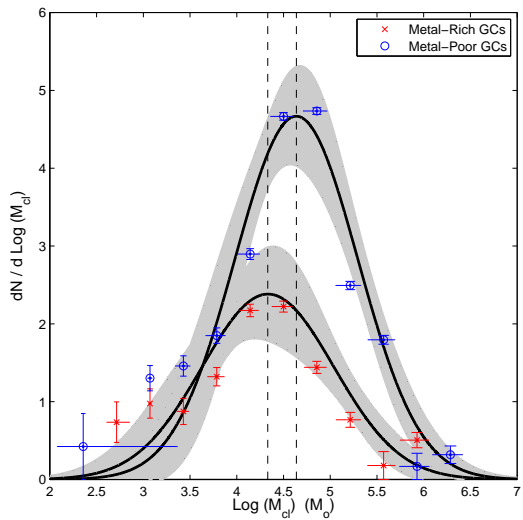}
\caption{Comparison of the PDMF of the metal--rich and of the metal--poor populations of Galactic GCs from the Harris catalogue.}
\label{fig:MR_vs_MP}
\end{figure}

We found different values of the turnoff mass of the PDMF from different works in the literature. In particular, we applied our approach to the data from the Harris catalogue \citep[][2010 version]{Harris96} with a mass--to--light ratio according to \cite{Bonatto12} and to the dynamical masses derived by \cite{Mandushev91}. We found a difference in the value of the PDMF turnoff equal to $\Delta M_{\mathrm{to,PDMF}} = 3.97\times 10^4 \; \mathrm{M}_\odot$ which translates to a difference $\Delta M_{\mathrm{to,IMF}} = 2.70 \times 10^4 \; \mathrm{M}_\odot$ in the turnoff of the IMF. This result mirrors the fact that a small difference in the IMF of GCs evolves to a difference of the same magnitude in the value of the turnoff of the present day observed MF. We note that the full \cite{Mandushev91} catalogue (147 clusters) has a turnover mass of $M_\mathrm{to} = 1.9\times10^5\;\mathrm{M}_\odot$, which compares well with other values from the literature, e.g. \cite{McLaughlin08} quote $M_\mathrm{to} = 1.2\times10^5\;\mathrm{M}_\odot$ for the full Galactic GC population. If we apply the \cite{Bonatto12} $M/L$ ratio method to the full Harris catalogue we get $M_\mathrm{to} = 3.10\times10^4\;\mathrm{M}_\odot$, which is lower. Thus we are more confident in the GCIMF predicted using the \cite{Mandushev91} dynamical masses, although the difference in the reconstructed mass function is small. This also highlights that the turnoff mass of the PDMF of metal--rich clusters is systematically lower than the full population and thus there is a difference in the PDMF of metal--rich and metal--poor globular clusters. In Figure \ref{fig:MR_vs_MP} we show the observed PDMF of the disc--bulge population and of the halo population of Galactic globular clusters. 
A lower turnover mass for a metal--rich subsample could be expected because these clusters are more likely to live where the potential is stronger, compared to metal-poor clusters which should be biased towards the halo.

Some initial assumptions of our model may influence the results. First of all, different values of the fraction of primordial binaries could result in different escape rates of stars from the cluster owing to two--body relaxation. However, \cite{Hurley01} showed that the inclusion of binaries has a little effect on the long--term evolution of a star cluster. As noted by \cite{Baumgardt03}, the initial concentration of the cluster affects its dissolution time. In order to estimate the effect of different values of the concentration, we ran $N$--body simulations of star clusters with a different initial configuration, concluding that in this case the dissolution time is affected by only a small percentage of its value (see also Figure \ref{fig:BM_tdiss}). In particular, we ran simulations of clusters initially under--filling their tidal limit. When considering the uncertainties of the model, according to our results the initial concentration of the clusters doesn't play a major role. This result reflects the fact that clusters with different initial concentration eventually relax to their maximum size, after which their evolution becomes remarkably self-similar. Another strong approximation is that for this preliminary study the external tidal field in \texttt{NBODY6} is reproduced by a static model that still includes some simplistic features, such as a point--mass bulge. It also doesn't take into account the effect of time--dependent features of the galactic potential. For the sample of objects that we are considering, the presence of a central rotating triaxial bar and a perturbation of the disc owing to a spiral pattern may influence significantly the dynamical evolution of the clusters. Furthermore, in our description we didn't include the effects of initial rotation on the cluster dynamics and on the evolutionary time scales. In fact, several works \citep[e.g.][]{Hung13} have been devoted to the study of rotating star clusters, concluding that initial rotation accelerates both the core collapse and cluster disruption, requiring a more detailed description of the evolutionary processes. Also, disc and bulge shocks are treated only in limited scope.

A more realistic prediction of the dynamical history of the galactic globular cluster would then require an improved description of the external tidal field, a problem that we aim to address in a forthcoming paper. In this improved potential we can evolve a range of star clusters, varying internal properties such as binary fraction and stellar density, to create a self--consistent and realistic description of cluster dissolution. We will also be able to look for differences (if any) in the IMFs of metal--poor and metal--rich clusters. Finally we note that, ideally speaking, a self consistent evolutionary model of star cluster systems should include both galaxy-- and cluster--scale internal evolution, combining the advantages of collisionless dynamics with the detail offered by \texttt{NBODY6}.

\subsection*{Acknowlegments}
This work was performed on the swinSTAR and gSTAR supercomputers at Swinburne University of Technology funded by Swinburne and the Australian Government's Education Investment Fund. We thank Guido Moyano Loyola and Chris Flynn for useful discussions and comments. LR acknowledges a CRS scholarship from Swinburne University of Technology.

\bibliography{biblio}

\bibliographystyle{mn2e}

\clearpage

\appendix

\section{Tidal limit in Generic Mass Distributions}
\label{sec:appendix}
     
\begin{figure}
\centering
\includegraphics[scale=0.5]{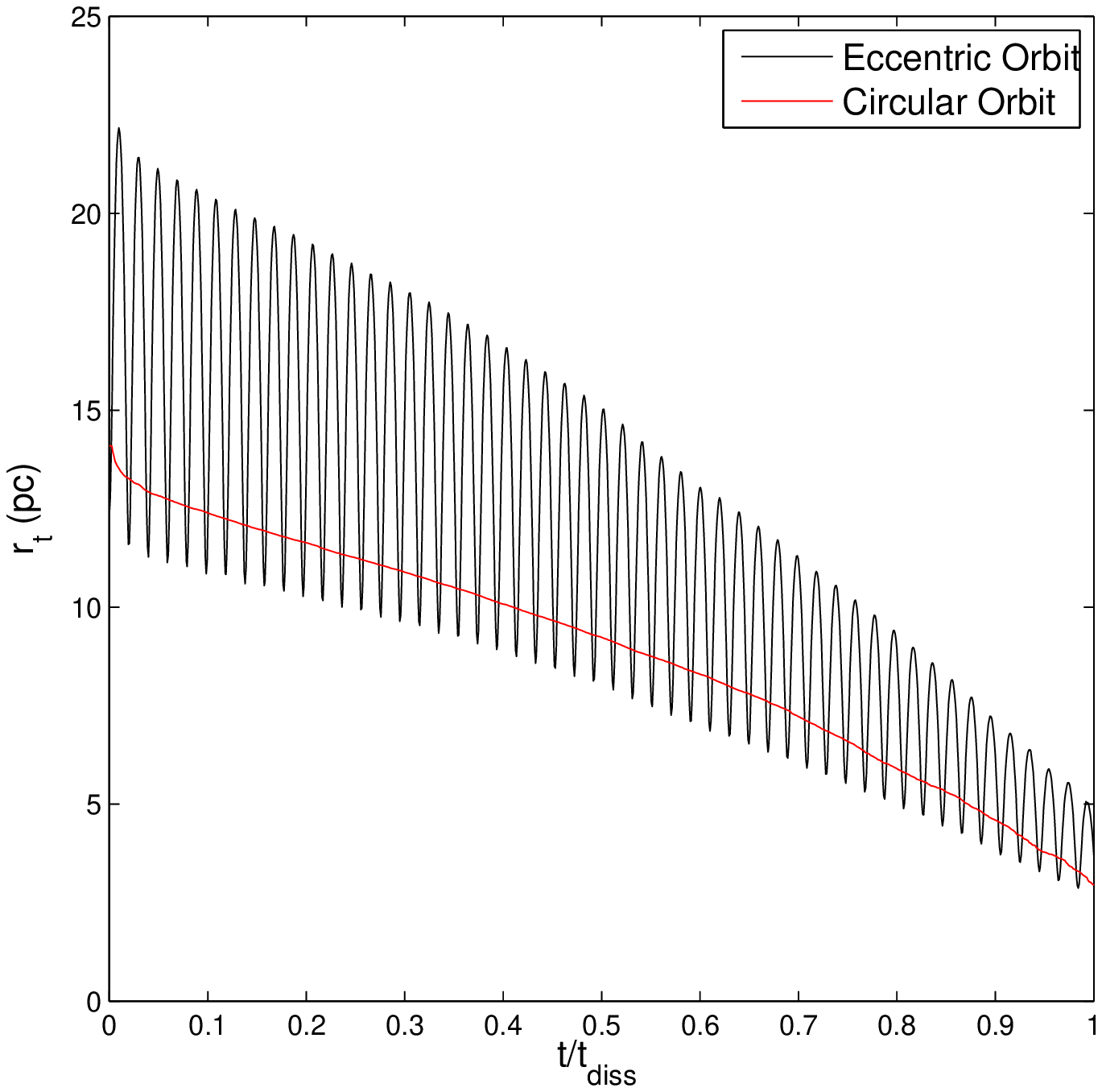}
\caption{Comparison between the evolution of the tidal radius for two clusters on an circular and on an eccentric orbit.}
\label{fig:rt_evol}
\end{figure}

According to \cite{King62}, the tidal limit of a cluster can be estimated as follows:
\begin{center}
``...As the cluster passes its perigalacticon, a star at a large distance from the cluster center will be detached by galactic tidal fields whereas a star at a small distance will not. We can then define the \textit{limit} as that point, on the line connecting the centre of the cluster with the galactic center, at which a star can remain on the line of centres with an acceleration along that line that is zero with respect to the cluster centre. That is, at the moment of perigalactic passage the star is pulled neither toward nor away from the cluster...''
\end{center}
Following the approach of \cite{King62} and under the assumption of an axisymmetric external potential
\begin{eqnarray}
\ddot{R_\mathrm{c}}&=&R_\mathrm{c}\dot{\theta}^2+f(R_\mathrm{c})\nonumber\\
				  &=&R_\mathrm{c}\dot{\theta}^2-\left.\dfrac{d\Phi(R)}{d R}\right|_{R_\mathrm{c}}
\end{eqnarray}
where $\Phi(R)$ is the total galactic potential, $R_\mathrm{c}$ is the Galactocentric distance of the cluster center and $\dot{\theta}$ is the angular velocity of the cluster.
The acceleration of a cluster star at the same time is given by
\begin{equation}
\ddot{R_\mathrm{s}}=R_\mathrm{s}\dot{\theta}^2-\left.\dfrac{d\Phi(R)}{d R}\right|_{R_\mathrm{s}}-\dfrac{GM_\mathrm{c}(R_\mathrm{s}-R_\mathrm{c})}{|R_\mathrm{s}-R_\mathrm{c}|^3}
\end{equation}
where $R_\mathrm{s}$ is the Galactocentric distance of a cluster star. The \textit{relative acceleration} is then
\begin{eqnarray}
\ddot{R_\mathrm{s}}-\ddot{R_\mathrm{c}}&=&(R_\mathrm{s}-R_\mathrm{c})\dot\theta^2-\left.\dfrac{d\Phi(R)}{d R}\right|_{R_\mathrm{s}}+\left.\dfrac{d\Phi(R)}{d R}\right|_{R_\mathrm{c}}-\nonumber\\
										& &-\dfrac{GM_\mathrm{c}(R_\mathrm{s}-R_\mathrm{c})}{|R_\mathrm{s}-R_\mathrm{c}|^3}\nonumber\\
									  &=&(R_\mathrm{s}-R_\mathrm{c})\left[\dot{\theta}^2-\dfrac{GM_\mathrm{c}}{|R_\mathrm{s}-R_\mathrm{c}|^3}\right] - \nonumber\\
									 & & -\left[\left.\dfrac{d\Phi(R)}{d R}\right|_{R_\mathrm{s}}-\left.\dfrac{d\Phi(R)}{d R}\right|_{R_\mathrm{c}}\right]\;\;\;.\nonumber
\end{eqnarray}
Since
\begin{equation*}
\dfrac{|R_\mathrm{s}-R_\mathrm{c}|}{R_\mathrm{c}}<< 1
\end{equation*}
we can rewrite the term in the second parenthesis in the rhs of the last equation as
\begin{equation}
\left[\left.\dfrac{d\Phi(R)}{d R}\right|_{R_\mathrm{s}}-\left.\dfrac{d\Phi(R)}{d R}\right|_{R_\mathrm{c}}\right]\simeq  \left.\dfrac{d^2\Phi(R)}{d R^2}\right|_{R_\mathrm{c}}(R_\mathrm{s}-R_\mathrm{c})
\end{equation}
and the relative acceleration as
\begin{equation}
\ddot{R_\mathrm{s}}-\ddot{R_\mathrm{c}}\simeq (R_\mathrm{s}-R_\mathrm{c})\left[\dot{\theta}^2-\dfrac{GM_\mathrm{c}}{       |R_\mathrm{s}-R_\mathrm{c}|^3}-\left.\dfrac{d^2\Phi(R)}{d R^2}\right|_{R_\mathrm{c}}\right].
\end{equation}
This will be zero when $(R_\mathrm{s}-R_\mathrm{c})$ has the magnitude $r_\mathrm{lim}$, given by
\begin{equation}
r_\mathrm{lim}^3=\dfrac{GM_\mathrm{c}}{\dot{\theta}^2-\left.\dfrac{d^2\Phi(R)}{d R^2}\right|_{R_\mathrm{c}}}
\label{eq:general_rt}
\end{equation}
This is a general expression for the tidal limit of a star cluster on the galactic plane, dependent on the form of the external axisymmetric potential and on the orbit of the cluster. In fact the term $\dot \theta$ is closely related to the $z$ component of the angular momentum $\dot \theta R^2 = h_\mathrm{z}$, representing an integral of motion in axisymmetric potentials. Note that in the case of circular orbits, the equation \ref{eq:general_rt} reduces to
\begin{equation}
r_\mathrm{lim}^3=\dfrac{GM_\mathrm{c}}{\dfrac{1}{R}\left.\dfrac{d \Phi(R)}{dR}\right| - \left.\dfrac{d^2\Phi(R)}{d R^2}\right|_{R_\mathrm{c}}}
\label{eq:circular_rt}
\end{equation} 
The relation usually adopted to predict the value of the tidal limit of a star cluster 
\begin{equation}
r_\mathrm{lim}^3= R_\mathrm{c}^3\left[ \dfrac{M_\mathrm{c}}{3M_\mathrm{g}}\right]
\end{equation}
is valid in the case of clusters orbiting on circular orbits within the gravitational potential generated by a point mass. The approximation for eccentric orbits
\begin{equation}
r_\mathrm{lim}^3= R_\mathrm{p}^3\left[ \dfrac{M_\mathrm{c}}{M_\mathrm{g}(3+e)}\right]
\end{equation}
where $R_\mathrm{p}$ is the perigalactic distance of the cluster, represents a compromise formula to predict the order of magnitude of the tidal limit a star cluster, but it can't predict its change as function of the phase. Figure \ref{fig:rt_evol} shows the evolution of the tidal radius of two clusters, one on a circular orbit at $R = 2 \; \mathrm{kpc} $ and the other on an eccentric orbit with $R_p = 2 \; \mathrm{kpc}$ and $e = 0.3$. 
Another example of this approach to predict the value of the tidal radius of clusters in different tidal fields can be found in \cite{Renaud11}.
 
\label{lastpage}
\end{document}